\documentclass[12pt]{article}

\usepackage{times}
\usepackage[dvipdfmx]{graphicx}
\usepackage{setspace}
\def\ltsima{$\; \buildrel < \over \sim \;$}
\def\ltsim{\lower.5ex\hbox{\ltsima}}
\def\lesssim{\lower.5ex\hbox{\ltsima}}
\def\gtsima{$\; \buildrel > \over \sim \;$}
\def\gtsim{\lower.5ex\hbox{\gtsima}}
\def\gtrsim{\lower.5ex\hbox{\gtsima}}

\begin{document}

\begin{center}
{\Large
Lethal radiation from nearby supernovae helps
to explain the small cosmological constant}

\vspace{1cm}

Tomonori Totani$^{1,2}$, Hidetoshi Omiya$^1$, Takahiro Sudoh$^1$, \\
Masakazu A.R. Kobayashi$^3$, Masahiro Nagashima$^4$ \\

\vspace{0.5cm}

$^1$Department of Astronomy, 
School of Science, The University of Tokyo, \\
7-3-1 Hongo, Bunkyo-ku, Tokyo 113-0033, Japan \\
$^2$Research Center for the Early Universe,
School of Science, The University of Tokyo,  \\
7-3-1 Hongo, Bunkyo-ku, Tokyo 113-0033, Japan\\
$^3$Faculty of Natural Sciences, National Institute of
  Technology, Kure College,  \\ 
2-2-11 Agaminami, Kure, Hiroshima 737-8506, Japan \\
$^4$Faculty of Education, Bunkyo University, \\ 
3337 Minami-Ogishima, Koshigaya-shi, Saitama 343-8511, Japan

\end{center}

\newpage

\begin{center}
{\bf \large Abstract}
\end{center}
The observed value $\Lambda_{\rm obs}$ of the cosmological constant
$\Lambda$ is extremely smaller than theoretical expectations, and the
anthropic argument has been proposed as a solution to this problem
because galaxies do not form when $\Lambda \gg \Lambda_{\rm
  obs}$. However, the contemporary galaxy formation theory predicts
that stars form even with a high value of $\Lambda / \Lambda_{\rm obs}
\sim$ 50, which makes the anthropic argument less persuasive. Here we
calculate the probability distribution of $\Lambda$ using a model of
cosmological galaxy formation, considering extinction of observers
caused by radiation from nearby supernovae. The life survival
probability decreases in a large $\Lambda$ universe because of higher
stellar density.  Using a reasonable rate of lethal supernovae, we
find that the mean expectation value of $\Lambda$ can be close to
$\Lambda_{\rm obs}$, and hence this effect may be essential to
understand the small but nonzero value of $\Lambda$. It is predicted
that we are located on the edge of habitable regions about stellar
density in the Galaxy, which may be tested by future exoplanet studies.

\section{Introduction}

The concordance cosmological model including a nonzero cosmological
constant, $\Lambda$, is accepted as the best description of the
universe observed (Weinberg et al. 2013; Bull et al. 2016).  However,
the observed value $\Lambda_{\rm obs}$ is smaller than theoretical
expectations by at least 50 orders of magnitude.  Furthermore,
$\Lambda$ works as the constant energy density of vacuum, and its
value is curiously similar to the decreasing matter energy density
just now in the long history of the universe.  The extreme smallness
and coincidence are perhaps the most difficult problem in physics
(Carroll 2001; Caldwell and Kamionkowski 2009).  Lack of persuasive
explanations has motivated to consider the anthropic argument,
assuming that $\Lambda$ is determined by a stochastic process in the
early universe. An observer will not appear in a universe of $\Lambda
\gg \Lambda_{\rm obs}$, because structure formation is suppressed
before galaxies form once the cosmic expansion is accelerated by
$\Lambda$ (Barrow and Tipler 1986; Weinberg 1987; Efstathiou 1995;
Martel et al. 1998; Garriga et al. 2000; Peacock 2007)\footnote{ See
  also Lineweaver \& Egan (2007) and Egan and Lineweaver (2008) for
  anthropic expectation about the ratio of matter to vacuum density of
  the universe, under the condition that the cosmological constant is
  the observed value (i.e., $\Lambda = \Lambda_{\rm obs}$).}.

A wide range of theoretical possibilities about stochastic
determination of $\Lambda$ have been discussed (e.g. Polchinski 2006;
Totani 2017).  Here we consider the simplest version in which
$\Lambda$ is the only random variable while other physical constants
and the total amount of matter in the universe are unchanged. 
  Though some theories predict that not only $\Lambda$ but also other
  physical constants may change, we leave these possibilities for
  future studies. If the prior probability distribution
  $dP_p/d\Lambda$ of $\Lambda$ extends down to $\Lambda = 0$ with a
  nonzero value, a natural distribution is flat per unit $\Lambda$
  (i.e. $dP_p/d\Lambda$ = const.)\footnote{If $\Lambda$ is bound
    to be positive, a flat distribution per unit $\ln \Lambda$ may
    also be possible without any characteristic scale. However, in
    this case we need to introduce a lower cut-off at $\Lambda \ll
    \Lambda_{\rm obs}$, otherwise the distribution is mathematically
    ill-defined. We do not know any low-energy physics leading to such
    a cut-off.}  around $\Lambda = 0$, because physically natural
  scales of $\Lambda$ are much larger than $\Lambda_{\rm obs}$. In
  this work we assume this prior distribution following previous
  studies (e.g., Efstathiou 1995; Martel et al. 1998; Garriga et
  al. 2000; Peacock 2007).  Then the probability distribution of
  $\Lambda$ for an observer, $dP_o/d\Lambda$, should be proportional
  to the number of observers per unit mass of cosmic matter,
  $n(\Lambda)$. A very small value of $|\Lambda| \ll \Lambda_{\rm
    obs}$ is statistically unlikely, and hence the coincidence is also
  explained.  Though $\Lambda$ can also be negative, the cosmic
  expansion would be pulled back to collapse before an age of $\sim$10
  Gyr when $\Lambda \ltsim -\Lambda_{\rm obs}$, allowing no observer
  like us in a universe with $\Lambda \ll -\Lambda_{\rm obs}$.  This
  means that inclusion of negative $\Lambda$ does not significantly
  affect the anthropic argument, and we consider the $\Lambda$
  distribution only in $\Lambda > 0$.

\begin{figure*}
\begin{center}
\includegraphics[width=120mm,angle=0]{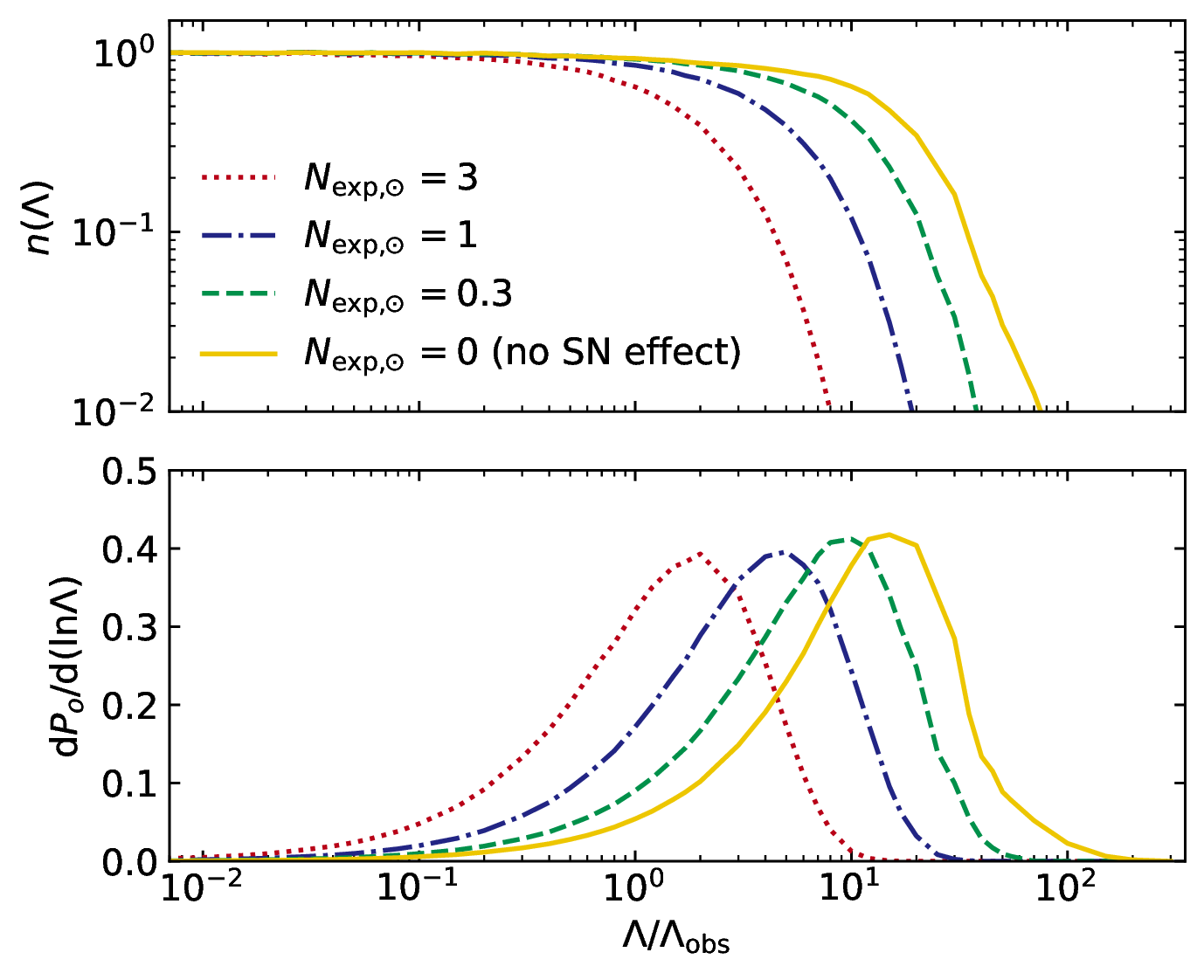}
\end{center}
\caption{Top panel: the number of observers $n(\Lambda)$ per unit mass
  of cosmic matter, normalized to unity in the limit of $\Lambda = 0$,
  for several different values of $N_{\exp, \odot}$ (the expected
  number of lethal events in the solar neighborhood during the time
  for evolution of life to an observer). When there is no supernova
  effect (i.e., $N_{\exp, \odot} = 0$), $n(\Lambda)$ is proportional
  to stellar mass produced in the universe up to a cosmic age of 15
  Gyr.  Bottom panel: the probability distribution of $\Lambda$ for an
  observer per unit $\ln \Lambda$, $dP_o/d(\ln \Lambda) = \Lambda \,
  dP_o/d\Lambda \propto \Lambda \, n(\Lambda)$, assuming a flat
prior distribution of $dP_p/d\Lambda =$ const.  }
\label{fig:Lambda_dist}
\end{figure*}

More quantitatively, $n(\Lambda)$ can be calculated based on the
theory of galaxy formation, if we assume that the number of observers
is proportional to stellar mass produced in the universe (Sudoh et
al. 2017; Barnes et al. 2018). The calculation of Sudoh et al. (2017)
by a semi-analytic model of cosmological galaxy formation (Nagashima
and Yoshii 2004) is shown by solid curves in Fig.
\ref{fig:Lambda_dist}, where $n(\Lambda)$ is calculated by stellar
mass\footnote{Sudoh et al. (2017) also considered a dependence of
  $n(\Lambda)$ on metallicity, but found that this effect is actually
  small (see their Fig. 4). This is because low metallicity galaxies
  are generally dwarf galaxies, and star formation in such galaxies is
  suppressed by supernova feedback, making their contribution to
  cosmic stellar mass budget small.}  produced up to a fixed age of
the universe (15 Gyr), and the cosmological parameters except for
$\Lambda$ are fixed to the 2015 determination by {\it Planck} (Planck
Collaboration 2016).  This model is based on the standard picture of
hierarchical structure formation driven by cold dark matter. Dark
matter halos are generated by the Monte-Carlo method, so that their
formation rate and merger history are consistent with the structure
formation theory. Then galaxies grow in dark halos by star formation,
which are calculated considering baryonic physics such as gas cooling,
supernova feedback, and galaxy mergers. Various observations,
including galaxy luminosity functions of local as well as
high-redshift galaxies, are broadly consistent with the predictions by
this model (Nagashima and Yoshii 2004; Kashikawa et al. 2006;
Kobayashi et al. 2007).  The predicted distribution of $dP_o/d(\ln
\Lambda)$ peaks at $\Lambda \sim 20 \, \Lambda_{\rm obs}$, and the
distribution extends to $\Lambda \sim 100 \, \Lambda_{\rm obs}$. The
probability of finding $\Lambda < \Lambda_{\rm obs}$ is only
$P_o(\Lambda < \Lambda_{\rm obs})$ = 6.7\%.  More recently,
  Barnes et al. (2018) found the distribution peak at
  $\Lambda/\Lambda_{\rm obs} = $50--60 and $P_o(\Lambda < \Lambda_{\rm
    obs})$ = 2\% using a numerical simulation of galaxy formation,
  which is roughly consistent with Sudoh et al. considering model
  uncertainties. This may imply that the value of $\Lambda$ is not
determined by the anthropic mechanism.

However, the amount of star formation is not the only factor to
determine the number of observers, but we should also consider planet
habitability around a star, which depends on locations and
environments in a galaxy (Gonzalez et al. 2001; Lineweaver et
al. 2004; Gowanlock et al. 2011; Forgan et al. 2017).  Radiation from
nearby supernovae is widely discussed as a possible cause of
extinction of terrestrial organisms on a planet, because high energy
radiation (gamma-rays and cosmic-rays) from a core collapse supernova
within $\sim$10 pc would have significant effects on the ozone layer
(Ruderman 1974; Whitten et al. 1976; Reid et al. 1978; Gehrels et
al. 2003).  Interestingly, the expected number of such nearby
supernovae in the solar neighborhood is of order unity during the time
span of terrestrial organisms ($\sim$ 0.5 Gyr) in the history of Earth
(Gehrels et al. 2003).  This coincidence implies that we may be living
on the edge of habitable regions concerning stellar density, and an
intelligent observer may not be allowed to appear in higher density
regions.  In this work we make a quantitative assessment of this
effect on the probability distribution of $\Lambda$.

\section{Stellar Densities in Galaxy Formation}

\begin{figure*}
\begin{center}
\includegraphics[width=120mm,angle=0]{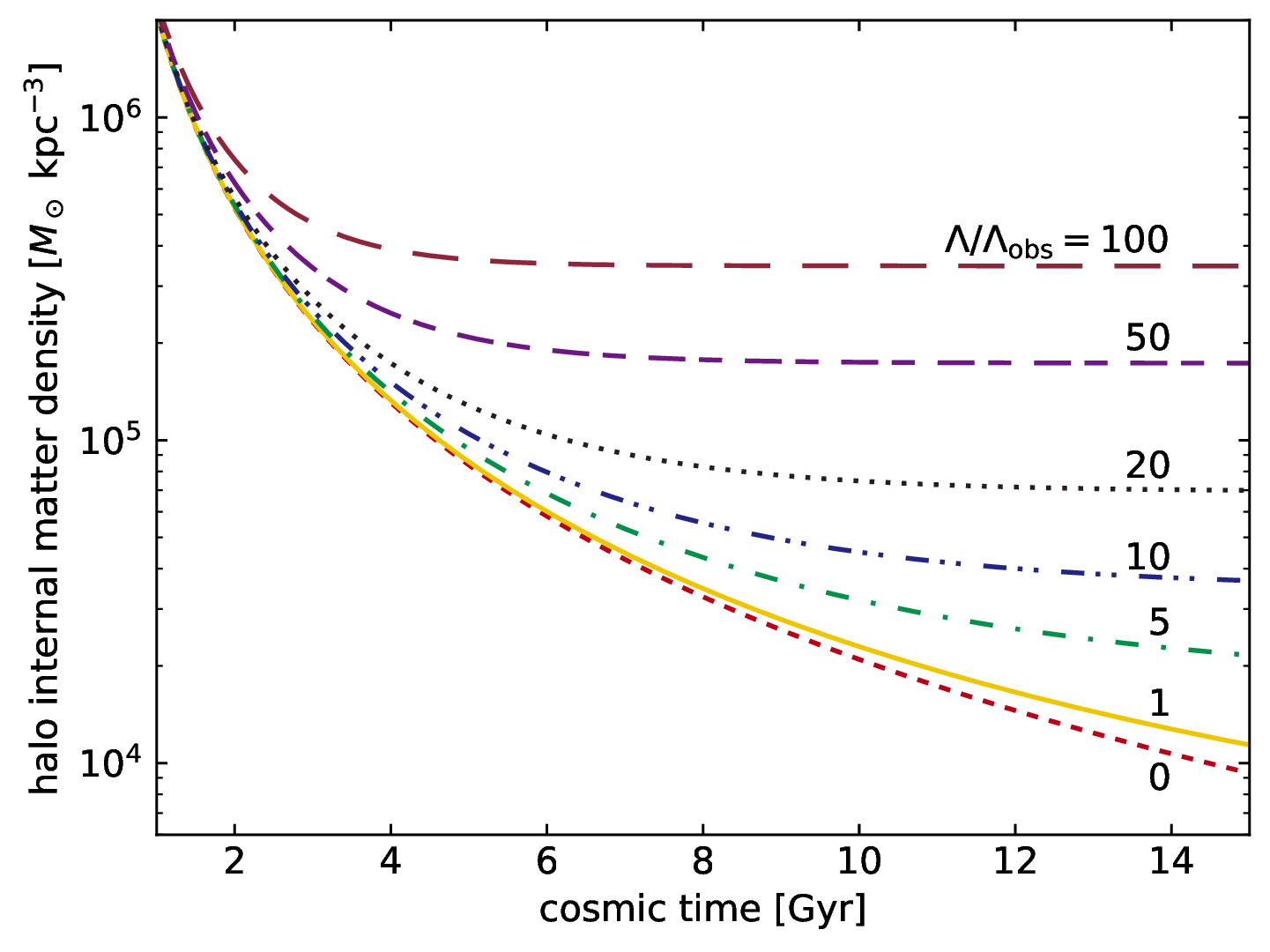}
\end{center}
\caption{Virial density $M_{\rm vir}/(4 \pi r_{\rm vir}^3/3)$
of a dark halo 
as a function of its collapsing time for 
different values of $\Lambda$, where $M_{\rm vir}$
and $r_{\rm vir}$ are the virial mass (baryon plus dark matter)
and virial radius, respectively. 
}
\label{fig:halodensity}
\end{figure*}

The key quantity in galaxy formation to determine stellar density is
the internal density of a gravitationally collapsed dark halo.  The
physics of dark halo formation can be understood by the analytic
spherical collapse model (see, e.g. Cooray and Sheth 2002 for a
review). The internal density (virial density) after the collapse,
$\rho_{\rm vir}$, does not depend on halo masses but it decreases with
cosmic time, as shown in Fig. \ref{fig:halodensity} for various values
of $\Lambda$.  When $\Lambda$ is large, decreasing energy density of
matter becomes lower than $\Lambda$ in an earlier epoch. After this
transition, $\rho_{\rm vir}$ becomes constant, and the number of
collapsing halos rapidly decreases.  Therefore, in a universe with
$\Lambda = 50 \, \Lambda_{\rm obs}$, internal density of any dark halo
is more than 10 times higher than that of a halo collapsing at a
cosmic time of 10 Gyr in the universe that we observe.  Stars are
expected to form only after the hot diffuse gas in a halo cools and
contracts, but the final size of a rotationally supported gas disk is
proportional to the virial radius of the host dark halo, because the
mean specific angular momentum given to halo gas is approximately
universal (Mo et al. 1998).  Therefore $\rho_{\rm vir}$ is a good
indicator of stellar density $\rho_*$ in a halo.

The Sun formed 9.2 Gyr after the big bang (a redshift of $z =
0.42$). According to the cosmic star formation history, about 90\% of
all stars formed by the present time (13.8 Gyr) are older than the Sun
(Borch et al. 2006), and hence in higher density regions.  This also
implies the effect of supernova radiation on life; if this effect is
not working, it is statistically more likely to find ourselves in
higher density regions than the solar neighborhood. It should also be
noted that such regions have more time for evolution of life.

For the first simple calculation of $n(\Lambda)$ with the nearby
supernova effect, we estimate stellar densities in the galaxy
formation model as follows.  Any model galaxy is either the central
galaxy in a dark halo or a satellite galaxy trapped by a dark halo as
a result of mergers with smaller halos.  The density of a stellar
population born in a central galaxy is assumed to be $\rho_* \propto
\rho_{\rm vir}$ of the host halo, while that in a satellite galaxy is
estimated from $\rho_{\rm vir}$ of the last host halo in which the
galaxy existed as the central galaxy. In some cases stellar density
would be changed by a galaxy-galaxy merger in a dark halo, but we
ignore this effect for simplicity.

The halo virial density of the Milky Way (MW) Galaxy is estimated as
$\rho_{\rm vir, MW} \sim 1.4 \times 10^4 \, M_\odot \, \rm kpc^{-3}$
(Klypin et al. 2002), which can be used as an indicator of $\rho_*$
around the solar neighborhood ($\rho_{*,\odot}$), assuming that the
Sun is a typical star formed around the main collapse epoch of the
present MW halo.  Then we can determine the stellar density relative
to the solar neighborhood value as $\rho_* / \rho_{*, \odot} =
\rho_{\rm vir}/\rho_{\rm vir, MW}$, for all stellar populations formed
at each time step in the galaxy formation model. It would be possible
to introduce more detailed modelings about internal structures of a
galaxy, but they depend on complicated baryon physics and model
uncertainties would become large. The primary cosmological effect of
$\Lambda$ to make stellar density higher has reasonably been taken
into account by this treatment.

\section{Probability Distribution of $\Lambda$
under the Nearby Supernova Effect}

Let $N_{\exp}$ for a stellar population be the expected number of
sterilizing events during the time for evolution of life to an
observer.  We introduce a parameter $N_{\exp, \odot}$, the value
around the solar neighborhood, which is rather uncertain but of order
unity if we consider nearby supernovae.  Obviously $N_{\exp}$ should
increase with stellar density, i.e., $N_{\exp} =
(\rho_*/\rho_{*,\odot}) \, N_{\exp, \odot}$.  Here, we implicitly
assumed that the supernova rate is the same for all stellar
populations, though the rate of core-collapse supernovae, which are
related to young stellar populations, should be small in old stellar
populations.  However, type Ia supernovae (SNe Ia) would also affect
habitability, which are related to white dwarfs and occur also in old
stellar populations (Totani et al. 2008). The SN Ia rate is about ten
times lower than the core-collapse rate in MW Galaxy, but they may
have more significant effect on habitability because of higher
luminosity (Gowanlock et al. 2011). Lethal events occurring in old
stellar populations are important in galaxies dominated by such
populations, like bulges or elliptical galaxies.

From the Poisson statistics, the probability of survival with no
lethal event is $P_s = \exp(- N_{\rm exp})$. Then $n(\Lambda)$ is
calculated by integrating all stellar mass produced in the universe,
but with the weight of $P_s$ for each stellar population in a galaxy.
Figure \ref{fig:Lambda_dist} shows $n(\Lambda)$ and the probability
distributions of $\Lambda$ for some different values of $N_{\exp,
  \odot}$.  The location of the distribution peak is reduced to
$\Lambda/\Lambda_{\rm obs} \sim 4$ when $N_{\exp, \odot} = 1$, and
$P_o(\Lambda < \Lambda_{\rm obs})$ is increased to 19\%. If $N_{\exp,
  \odot} = 3$, a rather small probability of $P_s(N_{\exp, \odot}) = $
5\% means that we are a lucky survivor, but the distribution peak is
further reduced to $\Lambda / \Lambda_{\rm obs} \sim 2$ with the
probability of $P_o(\Lambda < \Lambda_{\rm obs}) = $ 41\%.

\section{Discussion}

There are still many uncertain aspects about this calculation. The
critical distance to a lethal supernova suffers from uncertainties
about the effect of high energy particles on the Earth atmosphere, and
the degree of damage by ultraviolet light on organisms. Stellar
density should be calculated by simulations resolving internal
structure of galaxies in future studies, and supernova rates should
be calculated by local star formation history in a galaxy.  However,
the calculation shown here is based on reasonable parameters and
treatments, and hence lethal radiation from nearby supernovae must be
included in the anthropic consideration about $\Lambda$, which may be
the only way proposed so far to understand both the extreme smallness
and coincidence.  Supernovae played an important role for humankind to
discover the nonzero cosmological constant (Riess et al. 1998;
Perlmutter et al. 1999), but they may also be responsible for its
small value.

If there is any physical effect other than supernovae to reduce
habitability in high stellar density regions, it would also contribute
to making the expected value of $\Lambda_{\rm obs}$ smaller (e.g.,
comet bombardment by a field star passage or very wide binary star
systems affected by the Galactic potential, see Kaib 2018).
Especially, gamma-ray bursts (GRBs) have been discussed as a hazardous
astronomical event, whose impact may be comparable or even larger than
supernovae (Thorsett 1995; Melott et al. 2004; Piran and Jimenez
2014).  GRB rate is much lower than supernova rates, but their
luminosity is much brighter, and hence the maximum distance to a
lethal GRB is comparable to a typical size of a galaxy.  Piran et
al. (2016) proposed that lethal radiation from GRBs in nearby dwarf
galaxies prohibits an observer to exist in a low $\Lambda$ universe
($0 < \Lambda \ll \Lambda_{\rm obs}$), which would be important if the
prior $\Lambda$ distribution is weighted towards smaller values,
rather than being uniform per unit $\Lambda$.  It should be noted that
long-duration GRBs (LGRBs) are related only to young stellar
populations, but lethal events in old stellar populations are also
important to reduce $\Lambda_{\rm obs}$ by the effect considered in
our work.  Short-duration GRBs occur also in old stellar populations,
like SNe Ia, but their effect on habitability is likely much smaller
than LGRBs due to the smaller energy emitted per event. Another
important difference from supernovae is that LGRBs occur
preferentially in low metallicity regions. Since high stellar density
regions are generally metal rich in a galaxy, this should weaken the
effect of LGRBs on habitability in high density regions.

If the nearby supernova effect is indeed responsible for the observed
small value of $\Lambda$ as proposed here, a prediction is that we
should be located near the edge of habitable regions about stellar
density in a galaxy. This may be tested in future by development of
observational studies on exoplanets and their habitability.
Exoplanets in regions of higher stellar density would show smaller
probability of biomarker detection, even if they are apparently
habitable.

\section*{Acknowledgments}
This work was supported by JSPS/MEXT KAKENHI Grant Numbers 
15K05018, 17H06362, 17H02867, and 18K03692.

\newpage

\end{document}